\documentclass[11pt]{article}

\usepackage[final]{acl}

\usepackage{times}
\usepackage{latexsym}

\usepackage[T1]{fontenc}

\usepackage[utf8]{inputenc}

\usepackage{microtype}

\usepackage{inconsolata}

\usepackage{graphicx}

\usepackage{multirow}
\usepackage{makecell}
\usepackage{subcaption}
\usepackage{graphicx}
\usepackage{booktabs}
\usepackage{float}
\usepackage{longtable}

\usepackage{threeparttable}
\usepackage[dvipsnames]{xcolor}
\usepackage{fvextra}

\DefineVerbatimEnvironment{PromptVerbatim}{Verbatim}{
  fontsize=\fontsize{6.3}{6.1}\selectfont,
  breaklines=true,
  breakanywhere=true,
  breaksymbolleft={},
  breaksymbolright={},
  frame=single,
  framesep=1.5mm,
  framerule=0.4pt,
  xleftmargin=0pt,
  xrightmargin=0pt
}
\providecommand{\beginsupplement}{}

\usepackage{xspace}

\usepackage{amssymb}
\usepackage{pifont}
\newcommand{\cpmark}{ \textcolor{YellowOrange}{\ding{51}}  }
\newcommand{\cmark}{ \textcolor{green}{\ding{51}}  }
\newcommand{\xmark}{ \textcolor{red}{\ding{55}} }

\newcommand{\ours}{DeployBench\xspace}

\newcommand{\crchange}[1]{#1}

\title{\ours: Benchmarking LLM Agents for Research Artifact Deployment}

\author{
 \textbf{Yuanli Wang \textsuperscript{1}},
 \textbf{Yaoyao Qian \textsuperscript{2}},
 \textbf{Yue Zhang \textsuperscript{3}},
 \textbf{Hanhan Zhou \textsuperscript{4}},
 \\
 \textbf{Jindan Huang \textsuperscript{5}},
 \textbf{Tianfu Fu \textsuperscript{6}},
 \textbf{Qiuyang Mang \textsuperscript{7}},
 \textbf{Huanzhi Mao \textsuperscript{7}},
 \\
 \textbf{Wenhao Chai \textsuperscript{8}},
 \textbf{Wendong Fan \textsuperscript{9}},
 \textbf{Liqiang Jing \textsuperscript{3}}
\\
 \textsuperscript{1}Boston University,
 \textsuperscript{2}Northeastern University,
 \textsuperscript{3}University of Texas at Dallas,
 \\
 \textsuperscript{4}George Washington University,
 \textsuperscript{5}Tufts University, 
 \textsuperscript{6}xAI, 
 \\
 \textsuperscript{7}University of California, Berkeley, 
 \textsuperscript{8}Princeton University, 
 \textsuperscript{9}Eigent.Ai
\\
 \small{
   \textbf{Correspondence:} \href{mailto:yuanliw@bu.edu}{yuanliw@bu.edu}
 }
}

\begin{document}
\maketitle

\begin{abstract}
LLM agents have made rapid progress on software engineering and ML research tasks, but these advances often assume access to a working runnable environment. 
For research artifacts released alongside published papers, setting up such an environment from a fresh machine remains a major bottleneck. 
Existing environment setup benchmarks do not cover the full scope of research-artifact deployment, which involves multi-language toolchains, system-level dependencies beyond containers (e.g., GPU/CUDA and kernel configurations), and legacy artifact compatibility. 
We introduce \ours, a multi-domain benchmark of 51 research-artifact deployment tasks spanning AI/ML, computer systems, and scientific computing, covering all these dimensions.
Each task is verified by a hidden pipeline that executes the paper's designated experiment and checks its outputs. 
Evaluating \crchange{five} state-of-the-art LLMs with OpenHands yields pass-rates from 7.8\%--51.0\%.
Failures are dominated by a completion-judgment problem: \crchange{102} of \crchange{181} are agent-terminated self-stops, where the agent's pre-finish checks validate a different or weaker target than the paper-specific task requires.
\ours  highlights the gap between current agents and autonomous deployment, and offers a realistic testbed for scientific research agents.\footnote{\crchange{Repository: \url{https://github.com/pentium3/DeployBench}}}

\end{abstract}

\section{Introduction}

Large language model (LLM) agents have expanded beyond text generation into autonomous software engineering and scientific research, tackling repository-level software engineering tasks~\cite{Jimenez2023SWEbenchCL,Yang2024SWEagentAI}, reproducing open-source research code~\cite{Starace2025PaperBench,Yan2025LMRBench}, and performing ML engineering, data science tasks~\cite{Chan2024MLEbenchEM,Jing2024DSBenchHF} and frontier research~\cite{lyu2026mls,mang2025frontiercs}. However, these evaluations often assume access to a runnable environment where dependencies, build tools, and runtime stacks are already configured. In practice, research artifacts are not always released in a runnable state~\cite{nazario2025mitigating}, and the effort required to reproduce them can be substantial~\cite{mu2025understanding, Arora2025SetupBenchAS}. Setting up an artifact from scratch often requires resolving version-drifted dependencies, aligning GPU/CUDA stacks, patching legacy code against modern toolchains, and configuring system-level components such as custom Linux kernels that containers cannot host. This setup phase is often the primary barrier preventing researchers from reproducing, validating, and building upon published work.

Existing environment-setup benchmarks do not fully test this capability~\cite{Eliseeva2025EnvBenchAB,Xiao2025CSRBenchBL,Wang2026ResearchEnvBenchBA}. Prior work primarily evaluates general-purpose repository setup using container builds, static checks, or unit tests as success signals, or studies research-code deployment within narrower Python-based AI research or containerized settings. As a result, they do not measure whether agents can deploy diverse research artifacts from raw infrastructure and run a paper-specific target.

To address this gap, we introduce \ours, a benchmark designed to evaluate whether LLM agents can deploy published research artifacts from raw infrastructure to runnable environments. \ours includes 51 deployment tasks drawn from top-tier venues between 2008 and 2025, spanning AI/ML (25), computer systems (19), and scientific computing (7). The tasks cover 11 programming-language ecosystems and include 22 GPU-dependent workloads, 5 tasks that compile and boot custom Linux kernels in virtual machines, and 10 legacy artifacts (2011--2018) that often require compatibility repair against modern toolchains. Each task provides the agent with the paper, its code repository, and a fresh cloud VM with no pre-installed drivers. Task success is determined by a hidden 2-layer verifier (a global rule-based parser plus a task-specific check) that runs a lightweight version of the paper's designated experiment covering the main pipeline and key dependencies (e.g., training the model for one epoch), and checks expected outputs and runtime evidence.
We also confirm that all included tasks can be deployed by running the reference setup scripts.

We evaluate \crchange{5} state-of-the-art LLMs (GPT-5.3-Codex, \crchange{DeepSeek-V4-Pro,} Gemini-3.1-Pro, Grok-4.20, GPT-5.4-Mini) on \ours  with OpenHands as agent scaffold ~\cite{Wang2024OpenHandsAO}. The pass rates of these \crchange{5} LLMs range from 7.8\% to 51.0\%. Even the strongest model solves only about half the tasks. Failure analysis reveals five recurring root-cause patterns, with completion judgment as a dominant problem. Among \crchange{181} failures, \crchange{102} are agent-terminated self-stops. In these cases, the agent's own pre-finish checks often run against a different or weaker target than the paper-specific task requires. This suggests that agents still fall short both in deploying research artifacts and in deciding whether the resulting setup satisfies the paper-specific task.

\section{Related Work}

\paragraph{LLM agents for software engineering.}
LLM agents have been applied to a wide range of repository-level software engineering tasks~\citep{Luo2024RepoAgentAL,Chen2025CoReQAUP,Peng2025SWEQACL,Liu2024CodeUpdateArenaBK,Zhang2025CodeCriticBenchAH,Chan2024MLEbenchEM,Jing2024DSBenchHF}. To perform such tasks, agent scaffolds such as SWE-agent~\citep{Yang2024SWEagentAI} and OpenHands~\citep{Wang2024OpenHandsAO} have emerged, with progress measured on benchmarks such as SWE-bench~\citep{Jimenez2023SWEbenchCL} and Terminal-Bench~\citep{terminalbench}. However, before such downstream tasks can be performed, the repository must first be set up in a runnable environment, which remains a bottleneck due to version drift and dependencies that span Python packages, system libraries, GPU drivers, etc. 

\paragraph{Benchmarks for environment setup.}
Motivated by this gap, recent benchmarks have begun to study environment setup
and repository deployment directly. Installamatic~\citep{Milliken2024BeyondPI}
introduces a benchmark of repository installation tasks from open-source Python
projects, together with an agent that searches project documentation to install
and verify target repositories. ExecutionAgent~\citep{Bouzenia2024YouNI}
generates project-specific scripts for building arbitrary projects and running
their tests. EnvBench~\citep{Eliseeva2025EnvBenchAB} evaluates automated
environment setup across Python and JVM repositories, and
SetupBench~\citep{Arora2025SetupBenchAS} evaluates environment bootstrapping from
 Linux sandboxes with deterministic success commands. Repo2Run~\citep{Hu2025Repo2RunAB}
introduces an agentic framework and benchmark for building Docker-based
executable test environments for Python repositories.
EnConda-Bench~\citep{Kuang2026EnCondaBenchPT} is complementary to these
end-to-end setup benchmarks: it injects realistic README errors and provides
process-level trajectory evaluation of planning, error diagnosis, repair, and
final execution. Closer to research deployment,
CSR-Bench~\citep{Xiao2025CSRBenchBL} evaluates LLM agents on deploying computer
science research repositories, with tasks concentrated in NLP/CV/AI/ML/DM
topics, through setup-command generation, data preparation, and experiment
execution.
ResearchEnvBench~\citep{Wang2026ResearchEnvBenchBA}, the closest to our setting, benchmarks environment synthesis for research code, but operates inside Docker containers with CUDA drivers pre-installed, and covers only recent AI repositories in Python without older codebases or other domains.
Beyond benchmarks, agent systems such as HerAgent~\citep{Li2026HerAgentRT} and DockSmith~\citep{Zhang2026DockSmithSR} treat environment construction as a learning target, focusing on agentic strategy improvement rather than benchmark scope.

In contrast, \ours targets autonomous deployment in more realistic  settings: it starts from a bare metal Linux VM rather than a container, spans multiple research domains beyond AI, and involves legacy codebases that were written years ago and must now be made to build and run on modern systems. A task is solved only when the final environment executes a designated paper experiment.

\section{\ours}

We introduce \ours, a benchmark for evaluating whether LLM agents can construct environments that execute designated experiments from published research artifacts, starting from raw infrastructure. Beyond installing packages, this requires the agent to infer runtime requirements from the paper and repository, construct the needed software and system stack, repair compatibility issues, and determine whether the resulting environment is ready for the target experiment.

\subsection{Task Formulation}

\begin{figure*}[t]
\centering
\begin{minipage}[t]{0.62\linewidth}
\centering
\includegraphics[width=\linewidth]{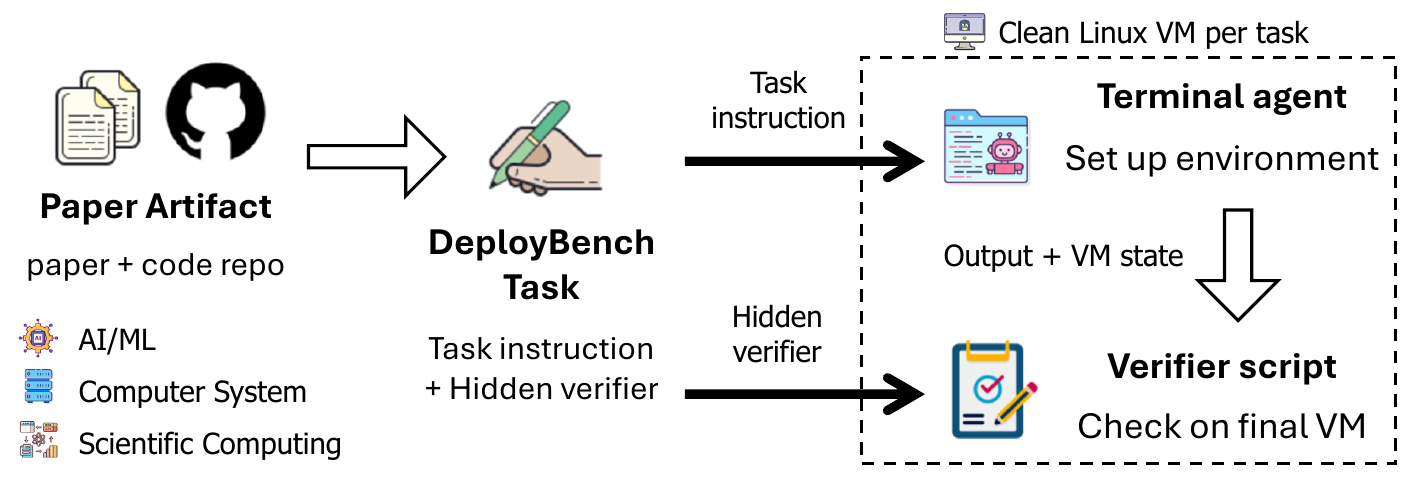}
\captionof{figure}{\ours overview.}
\label{fig:pipeline}
\end{minipage}
\hfill
\begin{minipage}[t]{0.35\linewidth}
\centering
\includegraphics[width=\linewidth]{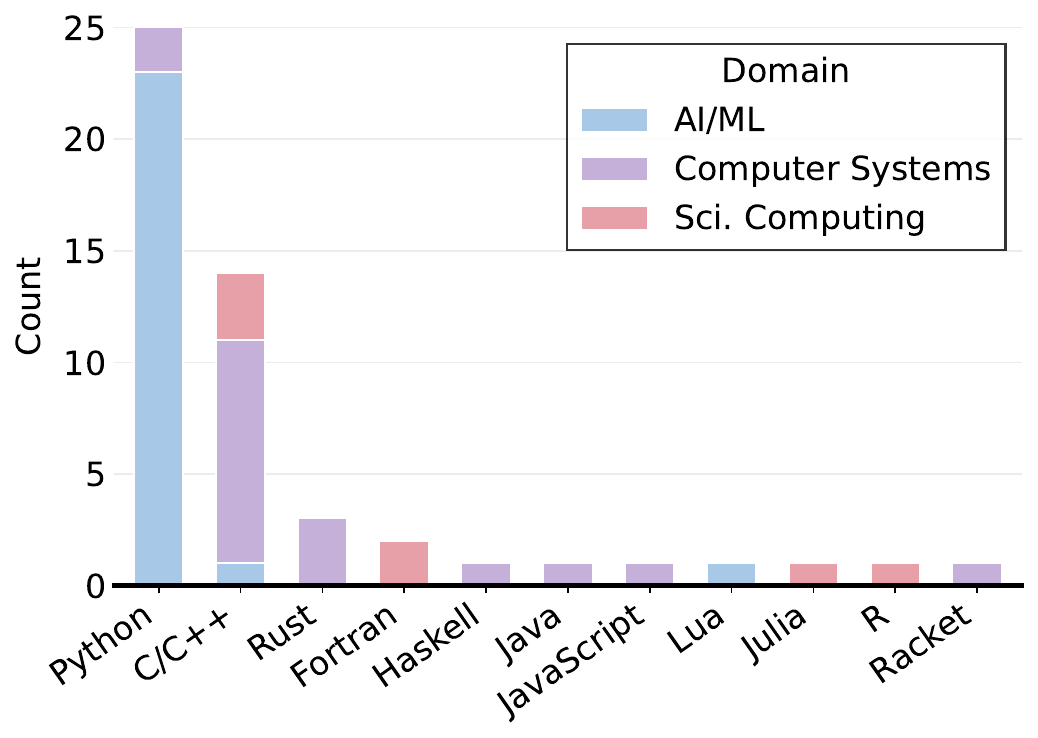}
\captionof{figure}{Distribution of tasks across domains and main languages.}
\label{fig:tasktypes}
\end{minipage}
\end{figure*}

Given a research paper and its code repository, a \ours task asks an agent to turn a clean cloud VM into an environment in which a designated experiment from the paper executes successfully (Figure~\ref{fig:pipeline}). 
Each task provides the agent with three inputs: the paper, the code repository, and a minimal Linux VM with no task-specific dependencies pre-installed. 
We disallow Docker so that success reflects native dependency resolution and system setup rather than reuse of pre-packaged containers. This is also necessary for system-level artifacts such as the  QEMU/kernel tasks.
Task success is determined by a hidden, task-specific verifier described in Section~\ref{sec:eval_method}.

\subsection{Benchmark Construction}
\label{sec:benchmark_construct}

\textbf{Paper Selection.}
\ours comprises 51 tasks drawn from research artifacts published between 2008 and 2025, spanning AI/ML (25 tasks from venues such as NeurIPS, ICML, ICLR, CVPR, ECCV, RSS, CoRL), computer systems (19 tasks from OSDI, SOSP, EuroSys, ATC, USENIX Security), and scientific computing (7 tasks from non-CS scientific domains, including JOSS and Cell artifacts). We select papers from top venues in each domain and prioritize artifacts with widely used public repositories. For systems and security tasks, we also prioritize papers that have passed artifact evaluation \cite{sysartifacts} when such badges or reports are available.
\crchange{We include an artifact only if its code and required assets are publicly available and the paper provides a reproducible experiment that can serve as a verification target.}
We manually inspect each candidate repository and attempt to run the artifact ourselves, excluding clearly broken, incomplete, or unreproducible artifacts. Each \ours task is validated by executing a verification target derived from a designated paper experiment on a fresh VM, rather than by import checks, command success, or repository-provided tests alone.

\textbf{Verifier Construction and Reproducibility Check.}
For each task, we first construct a working reference deployment on a clean VM to confirm that the artifact can be reproduced and to understand its dependencies and execution workflow. This step is used to validate task feasibility and verifier design. We then select a \emph{verification target} from the paper's original evaluation that is lightweight enough to run within a practical time budget but substantial enough to cover the main pipeline and its key dependencies (for example, training the model for one epoch). 
The task-specific verifier and reference setup script are manually constructed for each task. The verifier re-runs the target in the final agent-produced environment and checks the output against task-specific success conditions. As a reliability check, the reference setup script reproduces the environment from scratch and passes the verifier on all 51 tasks.
Per-task agent time budgets are set individually based on setup complexity and command execution time, ranging from 10 to 60 minutes.
We exclude candidates that cannot be reproduced during reference construction, such as those with unavailable datasets or model weights.

\textbf{Task Composition and Complexity.}
Across the 51 tasks, implementation languages span 11 ecosystems (Python, C/C++, Rust, Fortran, Julia, R, Haskell, Lua, Racket, JavaScript, Java), with several tasks combining multiple languages in a single build pipeline. The final set also spans a wide range of setup difficulty: 12 tasks are labeled \texttt{easy}, 20 \texttt{medium}, and 19 \texttt{hard}. These labels are assigned by three authors with graduate-level CS research experience who constructed the reference deployments, based on dependency layers, GPU/CUDA or QEMU/kernel setup, legacy compatibility repair, and debugging required during reference construction. Deployment requirements vary along multiple axes: 22 tasks require GPU execution while 29 are CPU-only, 5 require building a custom Linux kernel and booting it inside a QEMU virtual machine, and 10 are legacy artifacts released between 2011 and 2018 (not updated since 2020) that often require compatibility repair against modern OS versions, compilers, and library APIs. Many tasks also include runtime-consuming setup steps that affect the deployment budget, such as compiling large codebases from source, downloading required weights or datasets to expected paths, or patching legacy APIs. Figure~\ref{fig:tasktypes} summarizes the distribution across domains and main languages.

\subsection{Evaluation} 
\label{sec:eval_method}

\begin{figure*}[htbp]
  \centering
  \includegraphics[width=0.9\linewidth]{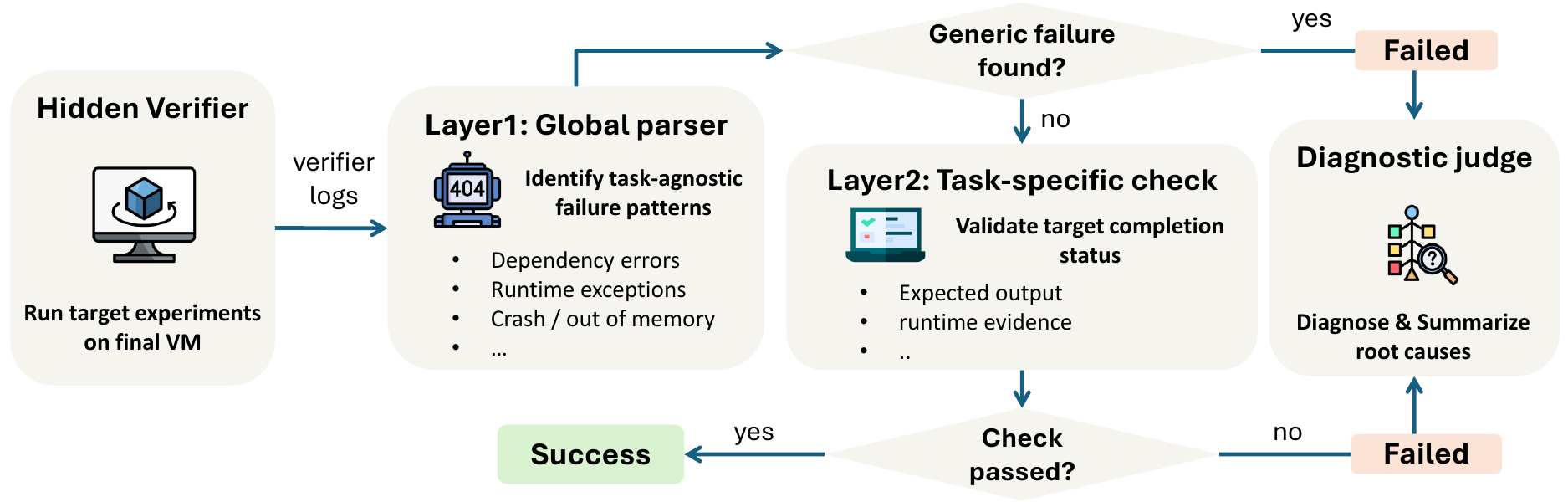}
  \caption{ Evaluation pipeline.} 
  \label{fig:eval1}
\end{figure*}

After the agent finishes, reaches the step limit, or times out, we evaluate the final VM state using a hidden task-specific verifier. The verifier has two layers (Figure~\ref{fig:eval1}) that check different failure cases. \textbf{Layer 1} is a global rule-based log parser that catches common execution failures shared across tasks, such as missing dependencies, compilation errors, runtime crashes, and out-of-memory errors. If no Layer-1 failure is detected, \textbf{Layer 2} runs task-specific checks for expected outputs, runtime evidence, generated artifacts, or service availability. Beyond Layer 1's generic failure checks, this second layer verifies whether the paper-specific target actually produces the required outputs or runtime evidence, rather than merely reaching a superficially runnable state.
A task is solved only if it passes both layers. For failed tasks, we additionally run a separate LLM-based diagnostic agent over the final VM state and verifier outputs to attribute the root cause (full prompt in Appendix~\ref{appendix:judge_prompt}). This diagnostic step is used only for failure attribution. All success rates are computed solely from deterministic verifier outcomes.

\crchange{\textbf{A representative verifier design example.}
For \texttt{adalora\_iclr23}~\cite{Zhang2023AdaLoRA}, the paper evaluates DeBERTa-v3-base on GLUE, and the proposed method consists of two core components: SVD-based adaptation and importance-aware rank allocation. The verifier runs the repository's own \texttt{run\_glue.py} on CoLA with both components enabled, reduced to 64 training examples and three optimizer steps. Success requires a crash-free GPU execution that produces new, non-empty training outputs during the verification run, including \texttt{rank\_pattern.json}, the output of the rank-allocation path. This run completes within minutes yet runs through the model and data loading stack, the SVD-based AdaLoRA modules, the optimization path, and adaptive rank allocation. Therefore, an environment that merely imports \texttt{loralib} or runs a toy example cannot pass (see the case study in Section~\ref{sec:case_study}).}

\crchange{\textbf{Avoiding false positives and false negatives.}
To reduce false positives, each verifier independently re-executes the designated experiment and requires task-specific runtime or output evidence that weak checks cannot produce; GPU tasks additionally verify CUDA availability.
To reduce false negatives, every verifier is validated by the independently constructed reference deployment on a fresh VM, and all 51 pass (Section~\ref{sec:benchmark_construct}).
In addition, the required workspace layout is specified in each task's instruction, and verifiers tolerate benign variation by checking alternative output locations.}

\subsection{Comparison with Existing Benchmarks}
\label{sec:comparison-existing-benchmarks}

Table~\ref{tab:benchmark-comparison} positions \ours against recent benchmarks for repository setup and environment synthesis along four axes: (1) multiple language ecosystems per task (Multi-Lang.), (2) per-task specific validation beyond generic success signals (imports, command success, builds, or repository-provided unit tests) (Per-Task Verifier), (3) host/VM-level setup beyond containers such as GPU/CUDA driver install, kernel builds, or QEMU execution (System-Level Setup), and (4) older artifacts requiring compatibility repair (Legacy Compat.).
Among benchmarks with task-specific validation, \ours uses a verifier that executes a lightweight target derived from a designated paper experiment on the final fresh-VM environment.
ResearchEnvBench~\cite{Wang2026ResearchEnvBenchBA} shares per-task verification and tests CUDA-stack alignment within Docker containers, but is restricted to AI repositories in Python with drivers pre-installed and excludes legacy artifacts. EnvBench~\cite{Eliseeva2025EnvBenchAB} covers only two language ecosystems (Python and Java). EnConda-Bench~\cite{Kuang2026EnCondaBenchPT} and CSR-Bench~\cite{Xiao2025CSRBenchBL} verify only via generic signals (Docker build , repository-provided unit tests , or generic command-execution signals) rather than task-specific validation. In contrast, \ours is the only benchmark combining all four axes.

\begin{table*}[t]
\centering
\small
\setlength{\tabcolsep}{4pt}
\resizebox{\textwidth}{!}{%
\begin{tabular}{llcccc}
\toprule
\textbf{Benchmark} &
\makecell{\textbf{Task Source}} &
\textbf{Multi-Lang.} &
\makecell{\textbf{Per-Task}\\\textbf{Verifier}} &
\makecell{\textbf{System-Level}\\\textbf{Setup}} &
\makecell{\textbf{Legacy}\\\textbf{Compat.}} \\
\midrule
EnvBench~\cite{Eliseeva2025EnvBenchAB}              & Python + Java OSS repos                          & \cpmark & \xmark  & \xmark & \xmark \\
SetupBench~\cite{Arora2025SetupBenchAS}             & OSS repos                          & \cmark  & \cmark  & \xmark & \xmark \\
Installamatic~\cite{Milliken2024BeyondPI}           & Python OSS repos                   & \xmark  & \xmark  & \xmark & \xmark \\
ExecutionAgent~\cite{Bouzenia2024YouNI}             & OSS repos                          & \cmark  & \xmark  & \xmark & \xmark \\
Repo2Run~\cite{Hu2025Repo2RunAB}                    & Python OSS repos                   & \xmark  & \xmark  & \xmark & \xmark \\
EnConda-Bench~\cite{Kuang2026EnCondaBenchPT}        & Python OSS repos + README errors          & \xmark  & \xmark  & \xmark & \xmark \\
ResearchEnvBench~\cite{Wang2026ResearchEnvBenchBA}  & AI research code repos             & \xmark  & \cmark  & \cpmark & \xmark \\
CSR-Bench~\cite{Xiao2025CSRBenchBL}                 & AI research code repos             & \xmark  & \xmark  & \xmark & \xmark \\
\midrule
\textbf{\ours (Ours)}                                & Multi-domain papers + code repos   & \cmark  & \cmark  & \cmark & \cmark \\
\bottomrule
\end{tabular}
}
\renewcommand{\arraystretch}{1.15}
\caption{Comparison of \ours with existing environment-setup benchmarks. \cmark, \cpmark, and \xmark~denote full, partial, and no support along each axis.}
\label{tab:benchmark-comparison}
\end{table*}

\section{Experiment}

\subsection{Experimental Setup}

\textbf{Models.} We evaluate \crchange{five} frontier language models under OpenHands agent scaffold: OpenAI GPT-5.3-Codex, OpenAI GPT-5.4-Mini, Google Gemini-3.1-Pro, xAI \crchange{Grok-4.20, and DeepSeek-V4-Pro.} 
\crchange{\ours itself is scaffold-agnostic, as verification depends only on the final VM state and the task-specific verifier. We fix a single scaffold so that all models are compared under identical tools, prompts, infrastructure, and budgets.}
For the diagnostic agent on failed runs, we use the Codex CLI tool (an OpenAI command-line agent based on GPT-5.3, high-reasoning mode) with SSH access to the final VM state and verifier logs.
\textbf{Prompting.} The same system prompt instructs the agent to set up the artifact natively on a fresh Linux VM (Docker disallowed), run a simple  test, and write a \texttt{RUNBOOK.MD} file summarizing the setup process. The full prompt is in Appendix~\ref{appendix:system_prompt}.
\textbf{Execution environment.} All runs use OpenHands~\cite{Wang2024OpenHandsAO} on a fresh Google Cloud VM running Ubuntu 22.04 (16 vCPUs, 64 GB RAM). GPU tasks use a VM with one NVIDIA L4 GPU (24 GB memory) and no pre-installed CUDA drivers. Each task uses a fresh VM, a task-specific time budget, and a fixed verification budget. The agent never sees the hidden verifier scripts.
\textbf{Metric.} Our primary metric is task success rate, the fraction of tasks whose verification pipeline returns a passing verdict.

\subsection{Model Performance}

\subsubsection{Overall Results}

\begin{table*}[htbp]
\centering
\small
\setlength{\tabcolsep}{3.5pt}
\begin{tabular}{llrrrrrr}
\toprule
Breakdown & Group & \# Tasks & GPT-5.3-Codex & \crchange{DeepSeek-V4-Pro} & Gemini-3.1-Pro & Grok-4.20 & GPT-5.4-Mini \\
\midrule
Overall & All tasks & 51 & 51.0\% & \crchange{47.1\%} & 27.5\% & 11.8\% & 7.8\% \\
\midrule
\multirow{3}{*}{Category}
& AI/ML & 25 & 52.0\% & \crchange{32.0\%} & 16.0\% & 8.0\% & 0.0\% \\
& Sci. Computing & 7 & 42.9\% & \crchange{57.1\%} & 57.1\% & 14.3\% & 0.0\% \\
& Systems & 19 & 52.6\% & \crchange{63.2\%} & 31.6\% & 15.8\% & 21.1\% \\
\midrule
\multirow{3}{*}{Difficulty}
& Easy & 12 & 66.7\% & \crchange{66.7\%} & 33.3\% & 25.0\% & 8.3\% \\
& Medium & 20 & 45.0\% & \crchange{40.0\%} & 40.0\% & 5.0\% & 10.0\% \\
& Hard & 19 & 47.4\% & \crchange{42.1\%} & 10.5\% & 10.5\% & 5.3\% \\
\midrule
\multirow{2}{*}{ \makecell{Repository \\ age }}
& \makecell{Legacy ($<$2020)} & 10 & 40.0\% & \crchange{30.0\%} & 20.0\% & 10.0\% & 0.0\% \\
& \makecell{Recent ($\geq$2020)} & 41 & 53.7\% & \crchange{51.2\%} & 29.3\% & 12.2\% & 9.8\% \\
\bottomrule
\end{tabular}
\caption{Task success rates by overall performance, task category, artifact difficulty, and repository age.}
\label{tab:breakdown-category-complexity}
\end{table*}

Table~\ref{tab:breakdown-category-complexity} reports the aggregate task success rates of all evaluated models on \ours: GPT-5.3-Codex leads at 51.0\%, followed by \crchange{DeepSeek-V4-Pro (47.1\%),} Gemini-3.1-Pro (27.5\%), Grok-4.20 (11.8\%), and GPT-5.4-Mini (7.8\%).
Even the strongest model solves only about half of the tasks, showing that end-to-end research artifact deployment remains a challenging task. The gap between models indicates that \ours\ can distinguish the capabilities of the model under the same agent scaffold.
\crchange{Note that the absolute scores and trajectory behaviors analyzed in Section~\ref{sec:further_analysis} reflect model-scaffold interactions under the fixed OpenHands scaffold.}

\subsubsection{Results by Task Category }

Table~\ref{tab:breakdown-category-complexity} also breaks performance down by task category, difficulty, and repository age. 
The differences between categories are mainly due to their deployment requirements. In AI/ML, 88.0\% of tasks require GPU-capable environments, making CUDA/GPU setup a major source of category-specific difficulty. In systems, the five QEMU/kernel tasks have an average success rate of only \crchange{24.0\%,} showing that VM- and kernel-level execution adds another source of deployment difficulty. Difficulty labels match the overall success pattern: at least one model solves 83.3\% of easy tasks, compared with \crchange{60.0\%} of medium tasks and 52.6\% of hard tasks.
Comparing legacy and recent tasks, since they run on the same modern OS image and hardware, this gap reflects compatibility drift rather than paper age itself: compatibility repair is needed in 5 of 10 legacy tasks versus 3 of 41 recent tasks, often because older code must be adapted to modern compilers and library APIs. Section~\ref{sec:case_study} gives a concrete example.

\subsubsection{Token Usage and Runtime Analysis}

\begin{table*}[htbp]
\centering
\small
\setlength{\tabcolsep}{4pt}
\begin{tabular}{lrrrrrrrrrr}
\toprule
& & & \multicolumn{5}{c}{Tokens (avg per task)} & \multicolumn{3}{c}{Runtime (minutes, avg per task)} \\
\cmidrule(lr){4-8} \cmidrule(lr){9-11}
Model & Success rate & Steps & Total & Input & Output & Solved & Failed & Total & Solved & Failed \\
\midrule
GPT-5.3-Codex  & 51.0\% & 44.5 & 2.15M & 2.12M & 30.7K & 1.63M & 2.69M & 18.1 & 14.6 & 21.8 \\
\crchange{DeepSeek-V4-Pro} & \crchange{47.1\%} & \crchange{58.7} & \crchange{2.11M} & \crchange{2.10M} & \crchange{16.3K} & \crchange{1.84M} & \crchange{2.36M} & \crchange{18.6} & \crchange{15.8} & \crchange{21.1} \\
Gemini-3.1-Pro & 27.5\% & 57.8 & 2.55M & 2.54M & 4.3K  & 1.49M & 2.94M & 15.9 & 10.1 & 18.1 \\
Grok-4.20      & 11.8\% & 15.2 & 0.75M & 0.75M & 4.1K  & 0.42M & 0.80M & 14.0 & 11.5 & 14.3 \\
GPT-5.4-Mini   &  7.8\% & 21.7 & 0.65M & 0.64M & 3.6K  & 0.45M & 0.66M &  6.1 & 10.2 &  5.8 \\
\bottomrule
\end{tabular}
\caption{Per-task token usage and runtime split by model and verifier outcome (Solved / Failed).}
\label{tab:cost-runtime}
\end{table*}

Table~\ref{tab:cost-runtime} reports per-model token usage and runtime. Token usage does not align with success ranking: Gemini-3.1-Pro consumes the most tokens (2.55M on average) yet ranks \crchange{only third} in success, while GPT-5.4-Mini uses the fewest tokens (0.65M) and has the lowest success rate. Within every model, failed runs use more tokens than solved runs (Codex 2.69M vs 1.63M, \crchange{DeepSeek 2.36M vs 1.84M,} Gemini 2.94M vs 1.49M, Grok 0.80M vs 0.42M, Mini 0.66M vs 0.45M), consistent with the higher Loop-style retry rate on failed runs (\crchange{56.9\%} vs \crchange{41.0\%,} Section~\ref{sec:trajectory_analysis}). Runtime follows the same per-model pattern except for GPT-5.4-Mini, which fails faster than it succeeds (5.8 min vs 10.2 min) because most of its failures are early self-stops.

\section{Further Analysis}
\label{sec:further_analysis}

To inform future agent design, we analyze the deployment-specific failure modes and trajectory patterns of the \crchange{five} evaluated models, drawing on verifier outputs, global-parser error types, and the diagnostic-agent reports described in Section~\ref{sec:eval_method}.

\subsection{Failure Analysis}
\label{sec:failure_analysis}

Table~\ref{tab:failure-layer} categorizes the \crchange{181} failed runs by where they are caught. The global check (Layer 1) detects \crchange{45.3\%} of runs.

The remaining \crchange{54.7\%}  pass the global check but still fail the task-specific verifier (Layer 2), meaning the environment reaches a superficially runnable state but does not produce the artifacts, services, outputs, or runtime evidence required by the designated experiment.

\begin{table}[htbp]
\centering
\small
\begin{tabular}{lrr}
\toprule
Failure location & Count & Percentage  \\
\midrule
Layer 1 -- Runtime crash & \crchange{45} & \crchange{24.9\%} \\
Layer 1 -- Dependency error & \crchange{32} & \crchange{17.7\%} \\
Layer 1 -- Compilation error & 5 & \crchange{2.8\%} \\
\midrule
Layer 1: global check  & \crchange{82} & \crchange{45.3\%} \\
Layer 2: task-specific check & \crchange{99} & \crchange{54.7\%} \\
\bottomrule
\end{tabular}
\caption{Failure locations in the evaluation pipeline.}
\label{tab:failure-layer}
\end{table}

From the diagnostic-agent reports, failures cluster into 5 recurring root-cause patterns:
\emph{dependency and package resolution} (incomplete or version-incompatible installs), \emph{GPU/CUDA setup} (CPU-only frameworks, missing drivers, mismatched CUDA-dependent libraries), \emph{native-build and toolchain} (failed compilation of non-Python components), \emph{artifact and output mismatch} (expected binaries, checkpoints, datasets, or outputs missing or misplaced), and \emph{system- and VM-level execution} (QEMU, kernel boot, service startup, or non-interactive configuration failures). 
The \crchange{19} \emph{all-fail} tasks (failed by all \crchange{five} models) typically combine multiple of these patterns: all \crchange{19} hit at least two and \crchange{16} hit three or more (\crchange{3.5} patterns per task on average).
The \crchange{31} \emph{mixed-outcome} tasks (at least one model succeeded and at least one failed) average only \crchange{2.6} patterns per task, indicating that all-fail tasks compound more failure modes simultaneously.
For example, \texttt{gan\_neurips14\_gpu} combines a wrong Theano version, an absent CUDA toolkit (\texttt{nvcc} and cuDNN missing), and a stale \texttt{pylearn2} commit.
\texttt{kflex\_sosp24} combines an incomplete QEMU rootfs (missing \texttt{libbenchmark}, \texttt{libjemalloc}, and \texttt{libabsl} shared objects) with a wrong system base.
Appendix~\ref{app:failure-modes} reports the full counts for each failure pattern.
These results show that per-task verification is necessary: generic Layer 1 checks would miss the \crchange{54.7\%} of failures caught only by the task-specific verifier.

\subsection{Trajectory Analysis}
\label{sec:trajectory_analysis}

We use trajectory analysis to identify which process signals are useful for agent design.
Overall, trajectory length is a weak success signal, while how models use the agent scaffold and how they recover from errors provide more useful  signals.

\paragraph{Trajectory length and phase coverage.}
Table~\ref{tab:traj-length} reports per-task average trajectory length under three units: agent steps, shell commands, and sub-commands (segments of a shell command split on top-level \texttt{\&\&}, \texttt{||}, or \texttt{;}  which expose work hidden inside chained commands. see Appendix~\ref{app:trajectory-phase}).
Longer trajectories do not imply better deployment.
\crchange{DeepSeek-V4-Pro and} Gemini-3.1-Pro \crchange{produce} the longest trajectories \crchange{(58.7 and 57.8 steps),} but \crchange{both underperform} GPT-5.3-Codex, which uses fewer steps (44.5).
Grok-4.20 issues only 10.4 commands on average, but these expand to 115.3 sub-command segments because it chains many operations inside each shell command.
Thus, trajectory length is useful for understanding how an agent works, but it does not predict whether the task will succeed.

We also classify sub-commands into deployment phases using heuristic keyword parsing (definitions in Appendix~\ref{app:trajectory-phase}).
Failed runs do not skip entire stages. Instead, compared with successful runs, they spend less time on environment inspection and more time in repeated install/build retries, suggesting that phase coverage alone is not enough to explain deployment success.
The full phase breakdown is in Appendix~\ref{app:trajectory-phase}.

\begin{table}[htbp]
\centering
\small
\begin{tabular}{lrrrr}
\toprule
Model & Steps & Cmds. &  \shortstack{Sub-\\cmds.}  &  \shortstack{Success \\ Rate} \\
\midrule
GPT-5.3-Codex & 44.5 & 37.2 & 178.7 & 51.0\% \\
\crchange{DeepSeek-V4-Pro} & \crchange{58.7} & \crchange{53.3} & \crchange{143.0} & \crchange{47.1\%} \\
GPT-5.4-Mini & 21.7 & 19.9 & 75.3 & 7.8\% \\
Gemini-3.1-Pro & 57.8 & 55.3 & 84.5 & 27.5\% \\
Grok-4.20 & 15.2 & 10.4 & 115.3 & 11.8\% \\
\bottomrule
\end{tabular}
\caption{Per-task average trajectory length by model.}  
\label{tab:traj-length}
\end{table}

\paragraph{Model-specific workflow and tool-use patterns.}
The models follow different workflow styles.
GPT-5.3-Codex follows a more plan-driven workflow, while Gemini-3.1-Pro produces longer Bash-heavy trajectories.
\crchange{DeepSeek-V4-Pro also produces long trajectories but makes the heaviest use of OpenHands' native file-read/file-edit actions (34/51 runs).}
GPT-5.4-Mini often self-stops with partial runbooks, and Grok-4.20 more often aborts after failed commands.
In terms of  tool-use patterns, 
one concrete difference is command-timeout use: GPT-5.3-Codex sets the OpenHands bash tool's optional \texttt{timeout} parameter on 100\% of bash calls, while Gemini-3.1-Pro sets it on only 2\%.
This difference explains why Gemini-3.1-Pro has many more soft-timeout events, where the shell tool stops waiting before the command completes.
These events make its trajectories look longer without reflecting additional deployment work.
Therefore, part of the trajectory-length gap reflects how models use the agent scaffold, not only how much deployment work they perform.
We do not claim that setting timeouts would directly improve success rates, but future agents should make agent tool-use policies explicit.
Details are discussed in Appendices~\ref{app:trajectory-patterns}.

\paragraph{Recovery after failed commands.}
Failed shell commands (non-zero exit or OpenHands soft-timeout) appear in both successful and failed runs. However, what differs is how the agent responds.
Using keyword and regex heuristics, we assign each failed command one of five recovery categories: \emph{Inspect} (read logs/files), \emph{Adapt} (change the command before retrying), \emph{Loop}, \emph{Abort/Finish}, or \emph{Other} (full definitions in Appendix~\ref{app:trajectory-recovery}).
Successful runs more often inspect logs or adapt the command (\emph{Inspect}+\emph{Adapt}: \crchange{52.9\%} vs.\ \crchange{36.3\%} in failed runs), while failed runs more often repeat waiting or retry loops (\crchange{56.9\%} vs.\ \crchange{41.0\%} in successful runs; Table~\ref{tab:recovery-by-outcome}).
Because OpenHands soft-timeouts are also counted as \emph{Loop}, especially for Gemini-3.1-Pro, this category captures unresolved waiting or retry behavior rather than only explicit repeated retries.
This suggests a concrete target for future agents: better log reading, dependency tracking, and retry decisions, rather than simply larger time or step budgets.

\begin{table*}[htbp]
\centering
\small
\begin{tabular}{lrrrrrr}
\toprule
Outcome & \shortstack{Failed Cmds.} & Inspect & Adapt & Loop &  \shortstack{Abort/Finish} & Other \\
\midrule
Success & \crchange{463} & \crchange{30.5\%} & \crchange{22.5\%} & \crchange{41.0\%} & \crchange{3.9\%} & \crchange{2.2\%} \\
Failed & \crchange{2{,}346} & \crchange{22.3\%} & \crchange{14.1\%} & \crchange{56.9\%} & \crchange{5.6\%} & \crchange{1.2\%} \\
\bottomrule
\end{tabular}
\caption{Recovery behavior after failed commands.}
\label{tab:recovery-by-outcome}
\end{table*}

\subsection{Completion Judgment}
\label{sec:completion_judgment}

\begin{table*}[htbp]
\centering
\small
\begin{tabular}{lr|rr|rr}
\toprule
Model &
\shortstack{Self-stop \\ failures} &
\shortstack{RUNBOOK-success \\ failures} &
\shortstack{RUNBOOK-partial \\ failures} &
\shortstack{Meaningful \\ check} &
\shortstack{Shallow-only \\ check} \\
\midrule
GPT-5.3-Codex  & 12/25 & 12/12 & 0/12  & 12/12 & 0/12 \\
\crchange{DeepSeek-V4-Pro} & \crchange{5/27} & \crchange{5/5}   & \crchange{0/5}   & \crchange{5/5}   & \crchange{0/5}  \\
GPT-5.4-Mini   & 42/47 & 17/42 & 20/42 & 33/42 & 9/42 \\
Gemini-3.1-Pro & 15/37 & 15/15 & 0/15  & 15/15 & 0/15 \\
Grok-4.20      & 28/45 & 11/28 & 12/28 & 21/28 & 7/28 \\
\bottomrule
\end{tabular}
\caption{Self-stop incidence over all failed runs, with RUNBOOK status and pre-finish check types over self-stops.}
\label{tab:stopping-behavior}
\end{table*}

Self-stop is the most common termination mode among failed runs: \crchange{102} of \crchange{181} failures are agent-terminated rather than reaching timeout or max-step. GPT-5.4-Mini (42/47) and Grok-4.20 (28/45) show the highest self-stop rates among the \crchange{five models, while DeepSeek-V4-Pro has the lowest self-stop rate (5/27) } (Table~\ref{tab:stopping-behavior}).

Within these \crchange{102} self-stopped failures, we find two types of completion-judgment errors.
First, at the RUNBOOK self-report level, \crchange{92} of \crchange{102} self-stops show RUNBOOK/verifier mismatch: \crchange{60} report \texttt{SUCCESS} while failing verification (\emph{RUNBOOK-success}), and 32 report \texttt{PARTIAL} while still failing verification (\emph{RUNBOOK-partial}). Only 1 of 33 self-stops with \texttt{PARTIAL}  reports ultimately passes verification.
Second, at the pre-finish check level, the check that the agent runs before finishing is often inadequate. All 12 of GPT-5.3-Codex's\crchange{,} all 15 of Gemini-3.1-Pro's\crchange{, and all 5 of DeepSeek-V4-Pro's} self-stopped failures do run a meaningful check (a target script, binary, or training/inference command on the artifact), but still fail verification because the agent validates the wrong target, runs a check that is weaker than the verifier, or misses task-specific requirements. We call this  \emph{self-validation drift}.
GPT-5.4-Mini (9/42) and Grok-4.20 (7/28) additionally show a subset of cases where the pre-finish check is \emph{shallow only} (imports, \texttt{--help}, toy examples, or RUNBOOK-only writes), never exercising the real artifact.

Future deployment agents need to derive a verifier-like completion check from the paper and repository: a lightweight run that covers the paper-specific target and key dependencies. Self-designed smoke tests and RUNBOOK self-assessment are not reliable substitutes.

\subsection{Case Studies}
\label{sec:case_study}

We present two representative cases: \texttt{adalora\_iclr23}~\cite{Zhang2023AdaLoRA} shows completion-judgment error. \texttt{silt\_sosp11}~\cite{Lim2011SILT} shows legacy compatibility repair requiring writing new compatibility code.

\subsubsection*{ \texttt{adalora\_iclr23} (with GPT-5.3-Codex)}
\textbf{Agent behavior.} The agent inspects the AdaLoRA repository and paper, creates a Python environment, installs a CUDA PyTorch stack and the local \texttt{loralib} package, and writes a custom smoke test around a toy \texttt{SVDLinear} model. The smoke test passes, after which the agent writes RUNBOOK = \texttt{SUCCESS} and self-stops. However, this smoke test exercises only  \texttt{loralib}  and does not require the HuggingFace evaluation stack used by the paper's experiment.
\textbf{Root cause.} The hidden verifier instead runs the paper's actual experiment through \texttt{run\_glue.py}: fine-tuning DeBERTa-v3-base on GLUE CoLA for a few steps with AdaLoRA's SVD-based LoRA modules. This requires the HuggingFace stack to be present in the environment, which the agent's toy-model smoke test never installed. The Layer 2 check aborts with error \texttt{ModuleNotFoundError: huggingface\_hub}.
\textbf{Takeaway.} The agent's smoke test checked  only \texttt{loralib}  rather than the lightweight paper-specific target used by the verifier. This illustrates \emph{self-validation drift}.

\subsubsection*{ \texttt{silt\_sosp11} (with GPT-5.4-Mini)}
\textbf{Agent behavior.} The agent inspects the SILT repository (a key-value store system built by the paper), installs build dependencies, and starts the build. Compilation fails on several legacy APIs, including old OpenSSL calls, removed TBB atomic types, and a private libc header. The agent patches the easy parts, such as include paths and simple type substitutions, but cannot solve the deeper TBB API gap. Finally, the agent self-stops with RUNBOOK = \texttt{PARTIAL}.
\textbf{Root cause.} A working deployment requires writing new compatibility code that wraps SILT's legacy TBB atomic API onto modern \texttt{std::atomic}, not just substituting types or includes. GPT-5.3-Codex succeeded by writing a 52-line \texttt{fawnds/tbb/atomic.h} shim and modifying eight other source files, and the resulting binary built and passed verification. GPT-5.4-Mini's text-level substitution did not provide this wrapper, and the agent stopped and honestly reported \texttt{PARTIAL}, rather than overclaiming success.
\textbf{Takeaway.} Legacy artifact deployment requires not just recognizing compatibility breakage but \emph{repairing it across the full dependency stack}, sometimes by writing new code (the TBB compatibility shim) rather than only editing existing source. Benchmarks restricted to actively-maintained repositories cannot cover this category (Table~\ref{tab:benchmark-comparison}).

\section{Conclusion}

We present \ours, a benchmark that evaluates whether LLM agents can deploy published research artifacts from raw infrastructure to a runnable state.
Across \crchange{five} frontier models, pass rates range from 7.8\% to 51.0\%.
Failure analysis reveals a dominant completion-judgment problem: \crchange{102} of \crchange{181} failures are agent-terminated self-stops.
These results point to three missing capabilities for agents: dependency repair, system-level setup, and strong completion checks.

\section*{Limitations}
The benchmark contains 51 tasks because each task requires manual reference deployment, a meaningful paper-specific target, a task-specific verifier, and a reproducible reference setup. This per-task cost limits scale compared with container-build benchmarks.
\crchange{\ours only includes artifacts whose code and required assets are publicly available and whose paper provides a reproducible experiment usable as a verification target.}
Our experiments fix the agent scaffold to OpenHands and vary only the underlying model.
\crchange{Thus, the absolute scores and trajectory behaviors reflect model–scaffold interactions and should not be generalized to all scaffolds.}
We leave additional agent scaffolds to future work.
More capable deployment agents may also make it easier to run unsafe or poorly maintained research code, so our evaluation runs agents in isolated cloud VMs.

\bibliography{sec/custom}

\appendix

\clearpage
\beginsupplement

\begin{center}
     \Large\textbf{Appendix}
\end{center}

\noindent The appendix is structured as follows:
\begin{itemize}
\setlength{\itemsep}{2pt}
\item Full list of benchmark source artifacts (Section~\ref{app:task-list}).
\item The full agent system prompt used in all runs (Section~\ref{appendix:system_prompt}).
\item The full diagnostic-agent prompt used to diagnose failed runs (Section~\ref{appendix:judge_prompt}).
\item Failure pattern counts across all-fail and mixed-outcome tasks (Section~\ref{app:failure-modes}).
\item Additional trajectory phase definitions and outcome-level phase distributions (Section~\ref{app:trajectory-phase}).
\item Additional model-specific trajectory signal definitions and metrics (Section~\ref{app:trajectory-patterns}).
\item Additional failed-command recovery definitions and distributions (Section~\ref{app:trajectory-recovery}).
\item Additional stopping behavior statistics by stop type (Section~\ref{app:stopping}).
\item \crchange{Disclosure of use of AI tools (Section~\ref{appendix:llm_use}).}
\end{itemize}

\section{List of Benchmark Source Artifacts}
\label{app:task-list}

Table~\ref{tab:task-list} lists every task in \ours by source artifact , venue and source year, grouped by the three domains used in Section~\ref{sec:benchmark_construct}.
Two source artifacts (Generative Adversarial Nets and Model-Agnostic Meta-Learning) each contribute two task variants, one CPU-only and one GPU-enabled, so the 51 tasks are drawn from 49 unique source artifacts.
\ours uses these public source artifacts for research evaluation of deployment agents by running lightweight targets derived from the corresponding paper experiments. The benchmark release provides task metadata and our setup/verifier scripts, and does not redistribute third-party repositories, datasets, or model weights.

\makeatletter
\if@twocolumn
  \onecolumn
  \def\restoretasklistcolumns{\twocolumn}
\else
  \clearpage
  \def\restoretasklistcolumns{\clearpage}
\fi
\makeatother

{\small
\begin{longtable}{p{0.75\textwidth}p{0.2\textwidth}}
\toprule
\textbf{Source Artifact / Paper Title} & \textbf{Venue / Source Year} \\
\midrule
\endfirsthead

\multicolumn{2}{l}{\small\textit{(Table~\ref{tab:task-list} continued from previous page)}} \\
\toprule
\textbf{Source Artifact / Paper Title} & \textbf{Venue / Source Year} \\
\midrule
\endhead

\midrule
\multicolumn{2}{r}{\small\textit{(continued on next page)}} \\
\endfoot

\bottomrule
\noalign{\vskip\abovecaptionskip}
\caption{Full list of source artifacts in \ours (51 tasks across 49 unique artifacts).}
\label{tab:task-list} \\
\endlastfoot

\multicolumn{2}{l}{\textit{AI/ML (25 tasks)}} \\
\midrule
Adaptive Budget Allocation for Parameter-Efficient Fine-Tuning & ICLR 2023 \\
ImageNet Classification with Deep Convolutional Neural Networks & NeurIPS 2012 \\
BERTScore: Evaluating Text Generation with BERT & ICLR 2020 \\
Counterfactual Reasoning for Out-of-distribution Multimodal Sentiment Analysis & ACM MM 2022 \\
Deep Mutual Learning & CVPR 2018 \\
ERGO: Event Relational Graph Transformer for Document-level Event Causality Identification & COLING 2022 \\
Generative Adversarial Nets (CPU variant) & NeurIPS 2014 \\
Generative Adversarial Nets (GPU variant) & NeurIPS 2014 \\
Grounding DINO: Marrying DINO with Grounded Pre-Training for Open-Set Object Detection & ECCV 2024 \\
Hyena Hierarchy: Towards Larger Convolutional Language Models & ICML 2023 \\
High-Resolution Image Synthesis with Latent Diffusion Models & CVPR 2022 \\
LightGCN: Simplifying and Powering Graph Convolution Network for Recommendation & SIGIR 2020 \\
Model-Agnostic Meta-Learning for Fast Adaptation of Deep Networks (CPU variant) & ICML 2017 \\
Model-Agnostic Meta-Learning for Fast Adaptation of Deep Networks (GPU variant) & ICML 2017 \\
MixText: Linguistically-Informed Interpolation of Hidden Space for Semi-Supervised Text Classification & ACL 2020 \\
MMSegmentation: OpenMMLab Semantic Segmentation Toolbox and Benchmark & GitHub release, 2024 \\
\crchange{Image Style Transfer Using Convolutional Neural Networks} & CVPR 2016 \\
Pixel Reasoner: Incentivizing Pixel Space Reasoning via Curiosity-Driven Reinforcement Learning & NeurIPS 2025 \\
PointNet++: Deep Hierarchical Feature Learning on Point Sets in a Metric Space & NeurIPS 2017 \\
QLoRA: Efficient Finetuning of Quantized LLMs & NeurIPS 2023 \\
Sample Efficient Grasp Learning Using Equivariant Models & RSS 2022 \\
One-shot Imitation Learning via Interaction Warping & CoRL 2023 \\
A Simple Framework for Contrastive Learning of Visual Representations & ICML 2020 \\
TimeMixer: Decomposable Multiscale Mixing for Time Series Forecasting & ICLR 2024 \\
YOLOv7: Trainable Bag-of-Freebies Sets New State-of-the-Art for Real-Time Object Detectors & CVPR 2023 \\
\midrule
\multicolumn{2}{l}{\textit{Computer Systems (19 tasks)}} \\
\midrule
Detecting Logic Bugs in Database Engines via Equivalent Expression Transformation & OSDI 2024 \\
Efficient Auditing of Event-driven Web Applications & EuroSys 2024 \\
Specification and Verification in the Field: Applying Formal Methods to BPF Just-in-Time Compilers in the Linux Kernel & OSDI 2020 \\
K9db: Privacy-Compliant Storage for Web Applications by Construction & OSDI 2023 \\
Fast, Flexible, and Practical Kernel Extensions & SOSP 2024 \\
NrOS: Effective Replication and Sharing in an Operating System & OSDI 2021 \\
Operating System Support for Safe and Efficient Auxiliary Execution & OSDI 2022 \\
PaSh: Light-touch Data-Parallel Shell Processing & EuroSys 2021 \\
Practically Correct, Just-in-Time Shell Script Parallelization & OSDI 2022 \\
Polyjuice: High-Performance Transactions via Learned Concurrency Control & OSDI 2021 \\
QSYM: A Practical Concolic Execution Engine Tailored for Hybrid Fuzzing & USENIX Security 2018 \\
RUDRA: Finding Memory Safety Bugs in Rust at the Ecosystem Scale & SOSP 2021 \\
Home, SafeHome: Smart Home Reliability with Visibility and Atomicity & EuroSys 2021 \\
Speedy Transactions in Multicore In-Memory Databases & SOSP 2013 \\
SILT: A Memory-Efficient, High-Performance Key-Value Store & SOSP 2011 \\
Storm: Refinement Types for Secure Web Applications & OSDI 2021 \\
Theseus: An Experiment in Operating System Structure and State Management & OSDI 2020 \\
$\mu$Slope: High Compression and Fast Search on Semi-Structured Logs & OSDI 2024 \\
Efficiently Mitigating Transient Execution Attacks using the Unmapped Speculation Contract & OSDI 2020 \\
\midrule
\multicolumn{2}{l}{\textit{Scientific Computing (7 tasks)}} \\
\midrule
GROMACS: High Performance Molecular Simulations through Multi-level Parallelism from Laptops to Supercomputers & SoftwareX 2015 \\
The Kaldi Speech Recognition Toolkit & ASRU 2011 \\
Oceananigans.jl: Fast and friendly geophysical fluid dynamics on GPUs & JOSS 2020 \\
OpenFOAM: Open Source CFD Toolbox & GitHub release, 2025 \\
QUANTUM ESPRESSO: A Modular and Open-Source Software Project for Quantum Simulations of Materials & J. Phys. Condens. Matter, 2009 \\
Integrated Analysis of Multimodal Single-Cell Data & Cell 2021 \\
wannier90: A Tool for Obtaining Maximally-Localised Wannier Functions & Comput. Phys. Commun., 2008 \\
\end{longtable}
}

\restoretasklistcolumns

\section{Agent System Prompt}
\label{appendix:system_prompt}

For every run we provide the agent with the same system prompt, reproduced verbatim in Figure~\ref{fig:agent_prompt}. The placeholder \texttt{<WORKDIR>} is substituted with the per-task absolute workspace path on the VM (e.g., \texttt{/home/vm/Desktop/<task>}). The agent is also given the per-task deployment instructions (paper PDF and code repository under \texttt{<WORKDIR>}), described separately in Section~\ref{sec:benchmark_construct}.

\begin{figure*}[p]
\centering
\begin{PromptVerbatim}
You are an autonomous LLM agent whose job is to set up a research paper's artifact on a provided Ubuntu machine and verify it with a simple test. You do NOT run the full paper experiments.

Hard constraints (must follow)
- You have terminal access to a real Ubuntu host with root privileges.
- Username: vm
- passwordless sudo inside the sandbox
- ALL project-related files (source tree, extracted archives, build outputs, virtualenvs/conda envs, downloaded datasets, logs, VM disk images, checkpoints) MUST live under:
  <WORKDIR>
- I will later run experiments from within <WORKDIR>, so your setup must be reproducible from there.
- Do NOT use Docker. If the authors provide a Dockerfile or "run via docker" instructions, ignore them and reproduce the same environment directly on the host.
- For system paths (e.g. /etc, /usr, /var): do NOT use FileEditor to view/edit files directly (it won't use sudo). Use execute_bash + sudo (or copy to a writable path, edit, then sudo-copy back).
- Only if the artifact is inherently system-level and cannot reasonably run directly (e.g., modified Linux kernel / kernel module requiring reboot, special kernel configs), you MAY create and use a VM inside the host and run it there (e.g., QEMU/KVM). If you do, keep all VM images/config under <WORKDIR> and provide exact VM run commands.

Goal
- Install dependencies, build/configure the artifact, and complete any required downloads (e.g. models, datasets) so that the environment is ready.
- Run a simple smoke test to confirm the setup works (e.g. a small demo). Do NOT run the full paper experiments or long-running benchmarks.
- Prefer the simplest native setup: system packages via apt, plus Python venv as needed.
- If the host machine has a GPU and the artifact can use it, you must make the GPU usable and use it; CPU-only fallback does not count as success.

Operating procedure (follow in order)
1) Initial inspection
   - List the contents of <WORKDIR>.
   - Identify: the code folder and the paper PDF.
   - Create vm in a new directory vm under <WORKDIR>, ONLY if you must use a VM

2) Read instructions and infer requirements
   - Read the paper PDF to understand: required OS/kernel assumptions, hardware assumptions, and what a minimal smoke test would be (not the full benchmarks).
   - Read README / INSTALL / scripts in the code.
   - If instructions assume Docker, translate them into native host steps.

Agent skills (optional -- use when helpful)
You have access to the following helper functions. Use them when they are useful; the file editor cannot read binary files (e.g. PDF). Run them via bash so output appears in the terminal. Use absolute paths (e.g. under <WORKDIR>).
- parse_pdf(file_path): Extract and print text from a PDF (e.g. the paper). Does not include figures.
- parse_docx(file_path): Extract and print text from a DOCX file.
- parse_pptx(file_path): Extract and print text from a PowerPoint file.
- parse_latex(file_path): Convert LaTeX to plain text and print.
Invoke from the OpenHands Python environment (required for imports):
  cd $HOME/OpenHands/openhands-cli && uv run python -c 'from openhands.runtime.plugins.agent_skills.file_reader.file_readers import parse_pdf; parse_pdf("ABS_PATH")'
Replace parse_pdf with parse_docx, parse_pptx, or parse_latex as needed, and ABS_PATH with the absolute file path. Output will appear in the command observation.

3) Dependency resolution
   - Determine all build/runtime dependencies (compilers, libraries, Python/Rust/Go/Java, CUDA, etc.).
   - Install with apt when appropriate.
   - For language-specific deps:
     - Python: Keep the paper's environment isolated from the agent's own runtime; Create and use a project-specific venv under <WORKDIR>/env/. Do not use uv, conda, or any other environment.
     - Rust/Go/Node: install toolchains if required
   - User consistency: Use sudo only for system packages (apt install) and global binaries (/usr/local/bin). Everything else -- language toolchains (Rust, Python venv), package managers, and build commands -- must be installed and run as the same non-root user I provided.
   - Record exact versions where possible (package versions, git commit hashes, pip freeze, compiler versions).

4) Build and configure
   - Build the artifact as required (e.g., make/cmake/bazel/meson).
   - Fix path issues so everything runs when invoked from within <WORKDIR>.
   - If the artifact or smoke test requires downloaded models, datasets, or weights: download them and ensure the smoke test can use them. Do not skip downloads needed for a minimal run.

5) Run a simple smoke test
   - Execute a minimal check that the setup works (e.g. small demo, or a short run with minimal data). Do NOT run the full paper experiments or long benchmarks.

6) If a VM is needed (only as last resort)
   - Explain why native host execution is infeasible.
   - Use QEMU/KVM if available; create VM disk under <WORKDIR>/vm/.
   - Provide:
     - VM OS image source and checksums if applicable
     - VM config (CPU/RAM/disk) and exact launch command(s)
     - How files are shared between host and VM (e.g., virtiofs/9p/scp) while keeping project files in <WORKDIR>
     - The exact commands inside the VM to build and run the smoke test

Communication requirements (what you must report at the end)
- As soon as you have finished the setup (success or blocked), you MUST call the finish tool with a final message. Do not stop after a bash or file action without concluding.
- You MUST write a RUNBOOK file on disk so it can be collected: create exactly one file named RUNBOOK.MD under <WORKDIR> (i.e. <WORKDIR>/RUNBOOK.MD) with the runbook content. The runbook must include:
  - Setup status: SUCCESS / PARTIAL / FAILED (and why)
  - Artifact root directory (exact path under <WORKDIR>)
  - Exact commands to: 1) activate the environment, 2) run the smoke test you used to verify the setup
  - Optionally: how one would run the main paper experiment(s) (you did not run them; document for reference only)
  - Expected output of the smoke test (and, if documented, of the main experiment)
  - Versions: OS, kernel, gcc/clang, python, pip freeze (or equivalent)
  - If VM used: VM location, launch command, login credentials, and how to run the smoke test inside it
- Your final (finish) message to the user should summarize the same information; the RUNBOOK.MD file is the canonical record that will be collected from the machine.
- If you are blocked (e.g. missing dependency, build error), still write RUNBOOK.MD with what was done and what failed, then call finish.

Safety / guardrails
- Do not exfiltrate data.
- Do not delete or move the user-provided archives and paper.
- Do not modify files outside <WORKDIR> except for system dependency installation (apt) that is necessary.
\end{PromptVerbatim}
\vspace{-4mm}
\caption{Agent system prompt. The task-specific paths are replaced by placeholders.}
\label{fig:agent_prompt}
\end{figure*}

\section{Diagnostic-Agent Prompt for Failed Runs}
\label{appendix:judge_prompt}

For each failed run, we run an LLM-based diagnostic agent. The agent is given SSH access to the final VM state, the path to the reference setup script for the task, and the verifier outputs, and is asked to identify the root cause of the failure. Diagnostic reports are used only for failure attribution and never override the verifier's pass/fail verdict (Section~\ref{sec:eval_method}). The full system prompt is reproduced verbatim in Figure~\ref{fig:judge_prompt}. Placeholders \texttt{<SANDBOX\_IP>}, \texttt{<SANDBOX\_KEY>}, \texttt{<WORKDIR>}, and \texttt{<TASK\_DEF\_PATH>} are substituted at run time with the target VM's IP, the SSH key path, the per-task workspace, and the per-task definition directory respectively.

\begin{figure*}[p]
\centering
\begin{PromptVerbatim}
You are an expert systems engineer tasked with diagnosing why a research software environment failed to set up correctly on a bare-metal Linux machine.
You have SSH access to the target machine. Perform a thorough, systematic diagnosis and produce a structured report.

## Hard constraints (must follow)
- You have terminal access to a bare-metal Ubuntu sandbox with root privileges.
- sandbox IP: <SANDBOX_IP>
- Username: vm
- ssh key file: <SANDBOX_KEY>
- passwordless sudo inside the sandbox
- ALL project-related files (source tree, extracted archives, build outputs, virtualenvs/conda envs, downloaded datasets, logs, VM disk images, checkpoints) are live under:
  <WORKDIR>

## Reference Setup Procedure
The intended environment setup steps for this task are documented in:
  <TASK_DEF_PATH>/setup_reference.sh

Read this file first (`ssh -i <SANDBOX_KEY> -o StrictHostKeyChecking=no -o UserKnownHostsFile=/dev/null vm@<SANDBOX_IP> cat <TASK_DEF_PATH>/setup_reference.sh`) to understand what the correct setup looks like, then use it as the ground truth when diagnosing what went wrong.

## Your Goal
Identify the root cause(s) of environment setup failure. Be specific: name the exact package, version conflict, missing driver, failed download, or compilation error. Do not speculate -- base every conclusion on evidence you collect from the machine.

## Diagnosis Checklist
Work through each section. Skip sections that are clearly irrelevant to the task.

### 1. GPU & CUDA (if verify script needs gpu)
- GPU model and driver version: `nvidia-smi`
- CUDA toolkit version: `nvcc --version`, `ls /usr/local/cuda*`
- cuDNN presence: `find /usr -name 'cudnn*.h' 2>/dev/null`
- Check if installed packages match the CUDA version (e.g. torch+cu121 vs CUDA 12.4)

### 2. System Dependencies
- OS and kernel: `uname -a`, `lsb_release -a`
- Key system libraries:
  `dpkg -l | grep -E 'libssl|libffi|zlib|libc6|build-essential|cmake|ninja|openmpi|libopenblas'`
- Missing shared libraries for any binary: `ldd <binary> 2>&1 | grep 'not found'`
- Check for version mismatches between system libs and what packages expect

### 3. Python Environment
- Python version: `python3 --version`
- Active venv / conda env: `which python`, `pip --version`
- Installed packages: `pip freeze`
- Failed installs: check `pip install` logs if available
- Conflicting dependencies: look for version pins that are mutually incompatible
- Missing packages that the code imports but are not installed

### 4. Dataset & Model Downloads
- Check if expected data files exist in the workspace
- Check download logs for HTTP errors, timeouts, or authentication failures
- Verify file integrity if checksums are available (truncated downloads)
- Check available disk space: `df -h`

### 5. Compilation & Build Errors
- Check for `.log` files in the build directory
- Look at `make` output if captured
- Check for missing headers or libraries that caused compilation to fail
- Verify compiler version: `gcc --version`, `g++ --version`, `nvcc --version`
- Check CMake cache for misconfigured paths

### 6. Version Compatibility
- Identify any API breakage between installed package versions and the code
- Check if the code was written for an older version of a framework (e.g. old PyTorch API, old TF 1.x style code)
- Look for deprecation errors or `AttributeError`/`ImportError` in logs

### 7. Verify Script Failure
- Show the exact output of the verify script: `cat verify.sh`, then run it and capture output
- Identify which specific check failed and why
- Show the content of any missing output files that were expected

### 8. Runtime Errors
- Check workspace for any `.log`, `stderr`, `stdout`, or `error` files
- Look at the last N lines of any training/evaluation logs
- Identify OOM (out-of-memory) errors, NCCL errors, or timeout issues

## Output Format

Produce a report with the following structure:

**Root Cause**: [One sentence summary of the primary failure reason]

**Evidence**:
- [Specific finding 1 with exact output or file path]
- [Specific finding 2]
- ...

**Secondary Issues** (if any):
- [Other problems found that may cause failure even after fixing root cause]

**Recommended Fix**:
- [Concrete, actionable steps to fix each issue]

Be concise. Paste only the relevant portions of command output, not full dumps. If a section shows no issues, say "OK" and move on.
\end{PromptVerbatim}
\vspace{-4mm}
\caption{Diagnostic-agent prompt for failed-run diagnosis. The runtime-specific values are replaced by placeholders.}
\label{fig:judge_prompt}
\end{figure*}

\section{Failure Pattern Counts for Tasks with Failed Runs}
\label{app:failure-modes}

This section provides the failure-pattern counts supporting the failure analysis in Section~\ref{sec:failure_analysis}. For each task, we map the \crchange{\emph{Root Cause} and \emph{Secondary Issues} sections of the diagnostic-agent reports} from failed runs to the five recurring failure patterns using keyword and regex heuristics. A task is counted under a pattern if any of its failed runs is attributed to that pattern.

Table~\ref{tab:app-failure-mode-incidence} reports the number of tasks in each group assigned to each pattern: the \crchange{19} all-fail tasks and the \crchange{31} mixed-outcome tasks (at least one model succeeded and at least one failed). Every all-fail task combines at least two patterns and \crchange{84\%} combine three or more, versus \crchange{42\%} on the mixed-outcome side. Dependency/package issues and native-build failures are the most frequent patterns in both groups, while GPU/CUDA setup, artifact/output mismatch, and system/VM-level execution are more concentrated in the all-fail set.

\begin{table}[htbp]
\centering
\small
\setlength{\tabcolsep}{4pt}
\begin{tabular}{p{0.37\linewidth}rr}
\toprule
Pattern & \makecell{All-fail \\ \crchange{(19)}} & \makecell{Mixed-outcome \\ \crchange{(31)}} \\
\midrule
Dependency / package / version          & 17 \crchange{(89\%)} & \crchange{24} (77\%) \\
Native build / toolchain                & \crchange{18 (95\%)} & \crchange{23 (74\%)} \\
GPU/CUDA setup                          & \crchange{11 (58\%)} & \crchange{12 (39\%)} \\
Artifact / output mismatch              & 11 \crchange{(58\%)} & \crchange{12 (39\%)} \\
System / VM-level execution             & \crchange{10 (53\%)} &  \crchange{9 (29\%)} \\
\midrule
Avg.\ patterns per task                 & \crchange{3.53}          & \crchange{2.58} \\
Tasks with $\geq 2$ patterns            & \crchange{19/19} (100\%) & \crchange{26/31 (84\%)} \\
Tasks with $\geq 3$ patterns            & \crchange{16/19 (84\%)}  & \crchange{13/31 (42\%)} \\
\bottomrule
\end{tabular}
\caption{Failure-pattern counts across the \crchange{19} all-fail tasks and the \crchange{31} mixed-outcome tasks. Per-task pattern union is computed across failed runs: a task is counted under a pattern if any failed run for that task is attributed to that pattern.}
\label{tab:app-failure-mode-incidence}
\end{table}

\section{Additional Trajectory Analysis}
\label{app:trajectory-phase}

We segment each shell command into \emph{sub-commands} by splitting on top-level \texttt{\&\&}, \texttt{||}, or \texttt{;}, which exposes work hidden inside large chained commands. This section provides the phase taxonomy used in Section~\ref{sec:trajectory_analysis}. Table~\ref{tab:app-phase-taxonomy} defines each sub-command phase, and Table~\ref{tab:app-phase-outcome} reports the phase distribution split by verifier outcome (success vs.\ failed).

\begin{table*}[htbp]
\centering
\small
\begin{tabular}{ll}
\toprule
Phase & Description \\
\midrule
Explore & Browse repository structure or metadata, such as listing files, searching paths, or checking git state. \\
Env. Inspect & Inspect the execution environment, such as OS, Python, GPU, driver, disk, or installed packages. \\
FileRead & Read documentation, source files, logs, or paper text through shell commands. \\
Install & Install system-level or language-level dependencies, such as apt, pip, npm, or cargo packages. \\
Build & Compile or build native components, binaries, kernels, or other generated artifacts. \\
FileEdit & Modify source files, scripts, patches, or configuration files through shell commands. \\
Config & Configure paths, environment variables, permissions, directories, symlinks, or runtime settings. \\
Assets & Download or clone external datasets, models, checkpoints, repositories, or other required assets. \\
Run/Test & Run artifact code, demos, tests, experiments, training, inference, binaries, or service checks. \\
Service & Start or manage services, background processes, emulators, simulators, or QEMU-based execution. \\
Finish & Write the required runbook or perform final completion-related actions. \\
\bottomrule
\end{tabular}
\caption{Deployment phase taxonomy used for sub-command-level trajectory analysis.}
\label{tab:app-phase-taxonomy}
\end{table*}

\begin{table*}[htbp]
\centering
\small
\resizebox{\textwidth}{!}{
\begin{tabular}{lrrrrrrrrrrr}
\toprule
Outcome & Explore & Env. Inspect & FileRead & Install & Build & FileEdit & Config & Assets & Run/Test & Service & Finish \\
\midrule
Success
& \crchange{52.8\%} & \crchange{9.9\%} & \crchange{12.7\%} & \crchange{3.6\%} & \crchange{3.4\%} & \crchange{1.8\%} & 5.6\% & 0.9\% & \crchange{5.5\%} & \crchange{2.4\%} & \crchange{1.4\%} \\
Failed
& \crchange{54.1\%} & \crchange{7.2\%} & \crchange{12.6\%} & \crchange{5.1\%} & \crchange{3.9\%} & \crchange{1.7\%} & \crchange{6.3\%} & \crchange{1.0\%} & \crchange{5.6\%} & \crchange{1.0\%} & \crchange{1.4\%} \\
\bottomrule
\end{tabular}
}
\caption{Sub-command-level deployment phase distribution, split by verifier outcome (successful vs.\ failed runs). Each cell reports the share of sub-commands labeled with that phase.}
\label{tab:app-phase-outcome}
\end{table*}

\section{Additional Trajectory Pattern Analysis}
\label{app:trajectory-patterns}

This section defines the trajectory signals used for the model-specific pattern analysis in Section~\ref{sec:trajectory_analysis}. Table~\ref{tab:app-trajectory-signal-defs} explains each signal, and Table~\ref{tab:app-trajectory-pattern-metrics} reports the corresponding model-level counts.
Here \emph{Soft timeout} is an observation produced when the shell tool stops waiting , not the bash tool's optional input parameter \texttt{timeout} chosen by the agent .

\begin{table*}[htbp]
\centering
\small
\begin{tabular}{p{0.20\linewidth}p{0.18\linewidth}p{0.52\linewidth}}
\toprule
Signal & Source & Definition \\
\midrule
Task tracker
& OpenHands action
& An OpenHands planning tool that lets the agent create or update an explicit TODO list during the run. We count whether each run uses this tool at least once. \\
\midrule
Structured File Actions
& OpenHands action
& Native OpenHands file-reading or file-editing actions. These are distinct from reading or editing files through Bash commands such as \texttt{cat}, \texttt{sed}, \texttt{tee}, or heredocs. \\
\midrule
MessageAction
& OpenHands action
& A natural-language message emitted by the agent. We use this as a signal of narration or status reporting, not as a deployment action by itself. \\
\midrule
Soft timeout
& OpenHands shell observation
& A tool-level observation marked with \texttt{exit\_code=-1}. It indicates that the shell tool stopped waiting before the command completed, rather than that the command necessarily terminated with a normal shell error. \\
\midrule
Max-step termination
& OpenHands runtime status
& A run stopped because OpenHands reached its configured maximum number of agent iterations. \\
\midrule
Heredoc use
& Derived command-level metric
& A run contains shell heredoc patterns, often used to write scripts, configuration files, or RUNBOOK content through Bash. \\
\bottomrule
\end{tabular}
\caption{Trajectory signals used in the model-specific pattern analysis. Some signals are native OpenHands actions or runtime observations, while others are derived during our post-processing.}
\label{tab:app-trajectory-signal-defs}
\end{table*}

\begin{table*}[htbp]
\centering
\small
\begin{tabular}{lrrrrrr}
\toprule
Model
& Task tracker
& \shortstack{Structured \\ File Actions}
& Heredoc
& \shortstack{Soft \\ Timeout}
& Max-step
& \shortstack{Message \\ Share} \\
\midrule
GPT-5.3-Codex  & 50/51 & 17/51 & \crchange{41/51} & 10/1898   & 4/51  & 0.1\%  \\
\crchange{DeepSeek-V4-Pro} & \crchange{18/51} & \crchange{34/51} & \crchange{15/51}  & \crchange{115/2719}  & \crchange{9/51}  & \crchange{0.4\%}  \\
GPT-5.4-Mini   & 2/51  & 0/51  & 46/51 & 47/1015   & 1/51  & 2.9\%  \\
Gemini-3.1-Pro & 0/51  & 12/51 & \crchange{27/51}  & 1299/2820 & 14/51 & 0.0\%  \\
Grok-4.20      & 4/51  & 24/51 & \crchange{25/51}  & 91/528    & 1/51  & 13.3\% \\
\bottomrule
\end{tabular}
\caption{Detailed model-specific trajectory metrics. \emph{Task tracker}, \emph{Structured File Actions}, and \emph{Heredoc} report the number of runs in which the behavior appears. \emph{Soft Timeout} is reported over paired shell-command observations. \emph{Max-step} is the number of runs that hit max-step termination. \emph{Message Share} is the fraction of OpenHands agent actions that are MessageActions. }
\label{tab:app-trajectory-pattern-metrics}
\end{table*}

\section{Additional Recovery Analysis}
\label{app:trajectory-recovery}

This section expands the recovery analysis in Section~\ref{sec:trajectory_analysis}. Table~\ref{tab:app-recovery-taxonomy} defines the five recovery labels, Table~\ref{tab:app-recovery-priority} shows how one label is selected when multiple behaviors appear, and Table~\ref{tab:app-recovery-full} reports the per-model distribution. The per-outcome distribution is reported in the main text (Table~\ref{tab:recovery-by-outcome}).

\textbf{Primary recovery label selection.}
When multiple recovery types apply, we pick the highest-priority match. Table~\ref{tab:app-recovery-priority} lists the nine ordered trigger conditions , each resolving to one of the five final labels.

\textbf{Per-model distribution.}
GPT-5.3-Codex actively diagnoses failures by inspecting logs (136/189), \crchange{and DeepSeek-V4-Pro shows a similar inspect-first profile (177/375);} Gemini-3.1-Pro mostly spins in \emph{Loop} (much of it soft-timeout polling rather than real recovery), and Grok-4.20 tends to give up early (Table~\ref{tab:app-recovery-full}). This supports the outcome-level finding (Section~\ref{sec:trajectory_analysis}) that successful runs are more often inspect or adapt , while failed runs are more often repeat waiting or retry loops .

\begin{table*}[htbp]
\centering
\small
\begin{tabular}{p{0.22\linewidth}p{0.68\linewidth}}
\toprule
Recovery label & Description \\
\midrule
Inspect
& The agent reads logs, error output, source files, documentation, or configuration files after a failed command. \\
\midrule
Adapt
& The agent changes the command, script, arguments, configuration, or environment before retrying. This includes generic modify-and-retry, switching to a different package, dependency version, toolchain, or installation path, and clean-and-rebuild actions. \\
\midrule
Loop
& The agent enters repeated or unresolved recovery behavior, including soft-timeout polling and repeated failure without meaningful change. \\
\midrule
Abort/Finish
& The agent stops, writes a final report, or otherwise gives up shortly after the failed command. \\
\midrule
Other
& The agent's next actions are an exact retry of the failed command, or do not fit any of the above categories. \\
\bottomrule
\end{tabular}
\caption{Recovery taxonomy used to classify responses after failed shell-command observations.}
\label{tab:app-recovery-taxonomy}
\end{table*}

\begin{table*}[htbp]
\centering
\small
\begin{tabular}{rp{0.20\linewidth}p{0.55\linewidth}}
\toprule
Priority & Final label & Trigger condition (first match wins) \\
\midrule
1 & Abort/Finish & No subsequent shell action, or the next action writes a runbook / matches a Finish signal. \\
2 & Loop & Soft-timeout polling: failed command has \texttt{exit\_code=-1} and the next action is empty input, Ctrl-C, or another soft timeout. \\
3 & Other & Next command is byte-equal (after whitespace normalization) to the failed command (exact retry without meaningful change). \\
4 & Inspect & Any of the next 1--3 actions reads logs, source, or documentation (\texttt{tail}, \texttt{cat}, \texttt{sed -n}, \texttt{grep}, \texttt{rg}, \texttt{less}, \texttt{find}, \texttt{ls}, \texttt{dmesg}, \texttt{pip show}). \\
5 & Adapt & Window text contains an install command together with version pins or driver/CUDA markers (\texttt{==}, \texttt{<=}, \texttt{>=}, \texttt{cuda}, \texttt{cu11}, \texttt{cu12}, \texttt{driver}, \texttt{headless}, \texttt{python[0-9]}). \\
6 & Adapt & Window text contains a cleanup command (\texttt{make clean}, \texttt{rm -rf}, \texttt{git clean}, \texttt{cargo clean}, \texttt{rebuild}). \\
7 & Adapt & Next command is different from the failed command and lands in an action phase (Install, Build, FileEdit-via-Bash, Config, Run/Test, Service). \\
8 & Loop & Repeated failure: subsequent window has further non-zero exits and contains at most two distinct normalized command forms. \\
9 & Other & None of the above conditions match. \\
\bottomrule
\end{tabular}
\caption{Priority order used to assign one final recovery label per failed command (higher priority wins when multiple conditions match). The same final label can arise from different trigger conditions: Loop from priorities 2 and 8, Adapt from priorities 5, 6, and 7, and Other from priorities 3 and 9.}
\label{tab:app-recovery-priority}
\end{table*}

\begin{table*}[htbp]
\centering
\small
\begin{tabular}{lrrrrrr}
\toprule
Model & Failed Cmds. & Inspect & Adapt & Loop & Abort/Finish & Other \\
\midrule
GPT-5.3-Codex
& 189 & 136 & 34 & 7 & 7 & 5 \\
\crchange{DeepSeek-V4-Pro}
& \crchange{375} & \crchange{177} & \crchange{73} & \crchange{111} & \crchange{7} & \crchange{7} \\
GPT-5.4-Mini
& 305 & 141 & 83 & 37 & 41 & 3 \\
Gemini-3.1-Pro
& 1748 & 181 & 229 & 1294 & 25 & 19 \\
Grok-4.20
& 192 & 28 & 15 & 76 & 70 & 3 \\
\bottomrule
\end{tabular}
\caption{Per-model recovery distribution after failed shell-command observations.}
\label{tab:app-recovery-full}
\end{table*}

\section{Additional Stopping Analysis}
\label{app:stopping}

This section expands  the stopping behavior analysis in Section~\ref{sec:completion_judgment}. We use three stop types in the trajectory analysis: \emph{self-stop} means that the agent explicitly ends the run, \emph{timeout} means that the run reaches its wall-clock time budget, and \emph{max-step} means that the run reaches the maximum number of allowed agent interaction steps. Table~\ref{tab:app-stop-type} reports verifier outcomes under each stop type.

\begin{table}[H]
\centering
\small
\begin{tabular}{llrr}
\toprule
Model & Stop type & Failed & Success \\
\midrule
GPT-5.3-Codex & Self-stop & 12/31 & 19/31 \\
GPT-5.3-Codex & Timeout & 9/16 & 7/16 \\
GPT-5.3-Codex & Max-step & 4/4 & 0/4 \\
\crchange{DeepSeek-V4-Pro} & \crchange{Self-stop} & \crchange{5/25} & \crchange{20/25} \\
\crchange{DeepSeek-V4-Pro} & \crchange{Timeout} & \crchange{15/17} & \crchange{2/17} \\
\crchange{DeepSeek-V4-Pro} & \crchange{Max-step} & \crchange{7/9} & \crchange{2/9} \\
GPT-5.4-Mini & Self-stop & 42/45 & 3/45 \\
GPT-5.4-Mini & Timeout & 4/5 & 1/5 \\
GPT-5.4-Mini & Max-step & 1/1 & 0/1 \\
Gemini-3.1-Pro & Self-stop & 15/28 & 13/28 \\
Gemini-3.1-Pro & Timeout & 8/9 & 1/9 \\
Gemini-3.1-Pro & Max-step & 14/14 & 0/14 \\
Grok-4.20 & Self-stop & 28/32 & 4/32 \\
Grok-4.20 & Timeout & 16/18 & 2/18 \\
Grok-4.20 & Max-step & 1/1 & 0/1 \\
\bottomrule
\end{tabular}
\caption{Verifier outcome by stop type. Each row uses runs with the corresponding stop type as the denominator.}
\label{tab:app-stop-type}
\end{table}

\section{\texorpdfstring{\crchange{Disclosure of Use of AI Tools}}{Disclosure of Use of AI Tools}}
\label{appendix:llm_use}

\crchange{We used Claude and ChatGPT as writing and coding aids for limited parts of the project, including language paraphrasing and script development. The benchmark design, task targets, verification criteria, analyses, and claims were determined and checked by the authors. Any LLM-assisted text or code was reviewed and revised by the authors before use. The LLMs and the diagnostic agent in our experiments are the objects of our evaluation, not writing or coding aids.}

\end{document}